\documentclass[twocolumn,superscriptaddress,nobibnotes,nofootinbib,floatfix,aps,prl,citeautoscript,reprint]{revtex4}

\usepackage{graphicx}
\usepackage{epsfig}
\usepackage{color}
\usepackage{latexsym}
\usepackage{amsmath,bm}
\usepackage{natbib}

\newcommand{\lsi}{Laboratoire des Solides Irradi\'es and ETSF, \'Ecole Polytechnique,
CNRS, CEA-DSM, 91128 Palaiseau, France}
\newcommand{\lpmcn}{LPMCN, Universit\'e Claude Bernard Lyon I and CNRS, 69622
Villeurbanne, France}
\newcommand{\llb}{Institut Rayonnement Mati\`ere de Saclay, Laboratoire L\'eon Brillouin,
CEA-CNRS UMR 12, CE-Saclay, F-91191 Gif-sur-Yvette, France}
\begin{document}

\title{Enhancing the superconducting transition temperature of BaSi$_2$ by structural tuning}

\author{Jos\'e A. Flores-Livas}
\affiliation{\lpmcn}
\author{R\'egis Debord}
\affiliation{\lpmcn}
\author{Silvana Botti$^{\ast}$}
\affiliation{\lsi}
\affiliation{\lpmcn}
\author{Alfonso San Miguel}
\affiliation{\lpmcn}
\author{Miguel A.L. Marques}
\affiliation{\lpmcn}
\author{St\'ephane Pailh\`es$^{\ast}$}
\affiliation{\lpmcn}
\affiliation{\llb}
\date{\today}

\begin{abstract}
We present a joint experimental and theoretical study of the
superconducting phase of the layered binary silicide BaSi$_2$.
Compared with the layered AlB$_2$ structure of graphite or
diboride-like superconductors, in the hexagonal structure of binary
silicides the $sp$$^3$ arrangement of silicon atoms leads to
corrugated sheets. Through a high-pressure synthesis procedure we are
able to modify the buckling of these sheets, obtaining the enhancement
of the superconducting transition temperature from 4\,K to 8.7\,K when
the silicon planes flatten out.  By performing {\it ab initio}
calculations based on density functional theory we explain how the
electronic and phononic properties of the system are strongly affected
by changes in the buckling.  This mechanism is likely present in other
intercalated layered superconductors, opening the way to the tuning of
superconductivity through the control of internal structural
parameters.
\end{abstract}

\maketitle


Nowadays, an important part of the activity research on
superconductivity is focused on intercalated layered crystal
structures. In these systems, where the relevant features for
superconductivity are intrinsic to the layers, the bond buckling of
the atoms forming the layers is a structural parameter known to damage
the superconducting properties, regardless of the nature of the
pairing mechanism~\cite{1}. The case of flat boron sheets in MgB$_2$
is well understood: superconductivity arises from a strong coupling
between the $sp$$^2$ $\sigma$-bonding intralayer electrons of boron
and the in-plane bond stretching phonons~\cite{2,3,4}. Besides, a
buckling of the boron honeycomb structure, as observed in
ReB$_2$~\cite{5}, was proved to decrease the superconducting
transition temperature $T_\textrm{c}$.  Nevertheless, recent studies
on graphite intercalated superconductors, namely (Yb,Ca)C$_6$
~\cite{6}, and the ternary silicide CaAlSi~\cite{7,8}, characterized
by large $T_\textrm{c}$s, point to the importance in these systems of
the electron-phonon (EP) coupling between the interlayer electrons and
the out-of-plane vibrational modes of the atoms composing the layer.
In that case, the buckling phonon modes corresponding to the antiphase
motion along the $c$-axis of the atoms in the sheets can lead to an
enhancement of the EP coupling. Experimentally, the effect of buckling
can either be explored by means of high-pressure or directly by
chemistry. An excellent testbed for such studies is the trigonal phase
of binary silicides, like CaSi$_2$ or BaSi$_2$, which constitutes a
family of layered intercalated superconductors closely related to the
graphite/diboride intercalated systems. Unlike the so-called
AlB$_2$-like structure of graphite/diboride compounds, where the
planes are flat, in the hexagonal structure of binary silicides the Si
planes buckle.

\begin{figure}[t]
  \centering \includegraphics[width=0.8\columnwidth]{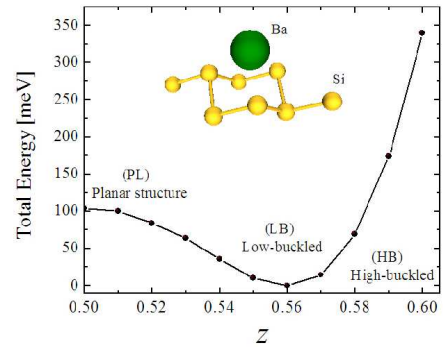}
  \caption{(Color online) Variation of the total energy of trigonal
    BaSi$_2$ per unit cell as a function of the buckling parameter
    $z$. The low-buckled regime $0.545 \lesssim z \lesssim 0.575$
    corresponds to values of $z$ that can be explored experimentally.}
  \label{fig:ENERGY}
\end{figure}

In this Letter we investigate superconductivity in BaSi$_2$ and, in
particular, we focus on its layered structure (EuGe$_2$-type
structure, $P$-3$m1$ space group)~\cite{9,10}. As it is sketched in
the inset of Fig.~\ref{fig:ENERGY}, trigonal BaSi$_2$ is made of
planes of Ba arranged in a triangular lattice, interspersed with
buckled hexagonal planes of Si. This phase is metallic and was found
to be superconducting with a critical temperature of
6.8\,K~\cite{11}. There are all the reasons to expect that BaSi$_2$ is
a standard $s$-wave superconductor, where the mechanism of
superconductivity can be obtained by studying the EP coupling.
However, recent theoretical studies~\cite{12} of trigonal BaSi$_2$
yielded a value of $T_\textrm{c}$ almost an order of magnitude smaller
than the early experimental finding of 6.8\,K. These studies were
based on state-of-the-art linear-response calculations within
density-functional theory (DFT), that have proved to describe very
accurately the superconducting properties of many other similar
materials.~\cite{13}. Analogous calculations for CaSi$_2$ in its
trigonal high pressure phase gave also a significant underestimation
of $T_{\textrm c}$~\cite{14}.  Interestingly, the application of
pressure on trigonal CaSi$_2$ was shown to increase the
superconducting temperature up to 14\,K~\cite{15}.  This finding was
discussed in terms of a structural phase transitions under pressure,
which is expected to soften the phonon modes associated to the
collapse of the corrugated Si planes~\cite{16}.

In order to elucidate the origin of superconductivity in binary
silicides, and at one time to shed light on the discrepant results
found in literature, we performed a joint experimental and {\it ab
  initio} study of the superconducting phase of BaSi$_2$.  Our
experimental data show that $T_\textrm{c}$ varies strongly with the
degree of buckling of the Si planes.  As in the case of
CaSi$_2$~\cite{15,16}, by flattening the Si planes, we increase the
superconducting transition temperature $T_\textrm{c}$ from 4\,K up to
8.7\,K.  Our extensive DFT calculations allow us to determine how the
buckling of the Si planes impacts the electronic and phonon band
structures, and finally the EP coupling.  Through the comparison of
experimental and theoretical results we infer that a soft phonon mode
corresponding to the out-of-plane Si vibrations, and very sensitive to
buckling, is responsible for the relatively high value of $T_{\textrm
  c}$ in BaSi$_2$ when Si planes get flatter.  Moreover, our
calculations put in evidence a correlation between the softening of
this phonon mode and the Si-Si bond buckling.  We expect that such a
behavior is a general property of conventional superconductors
presenting corrugated layers.


\begin{figure}[t]
  \centering \includegraphics[width=1.05\columnwidth]{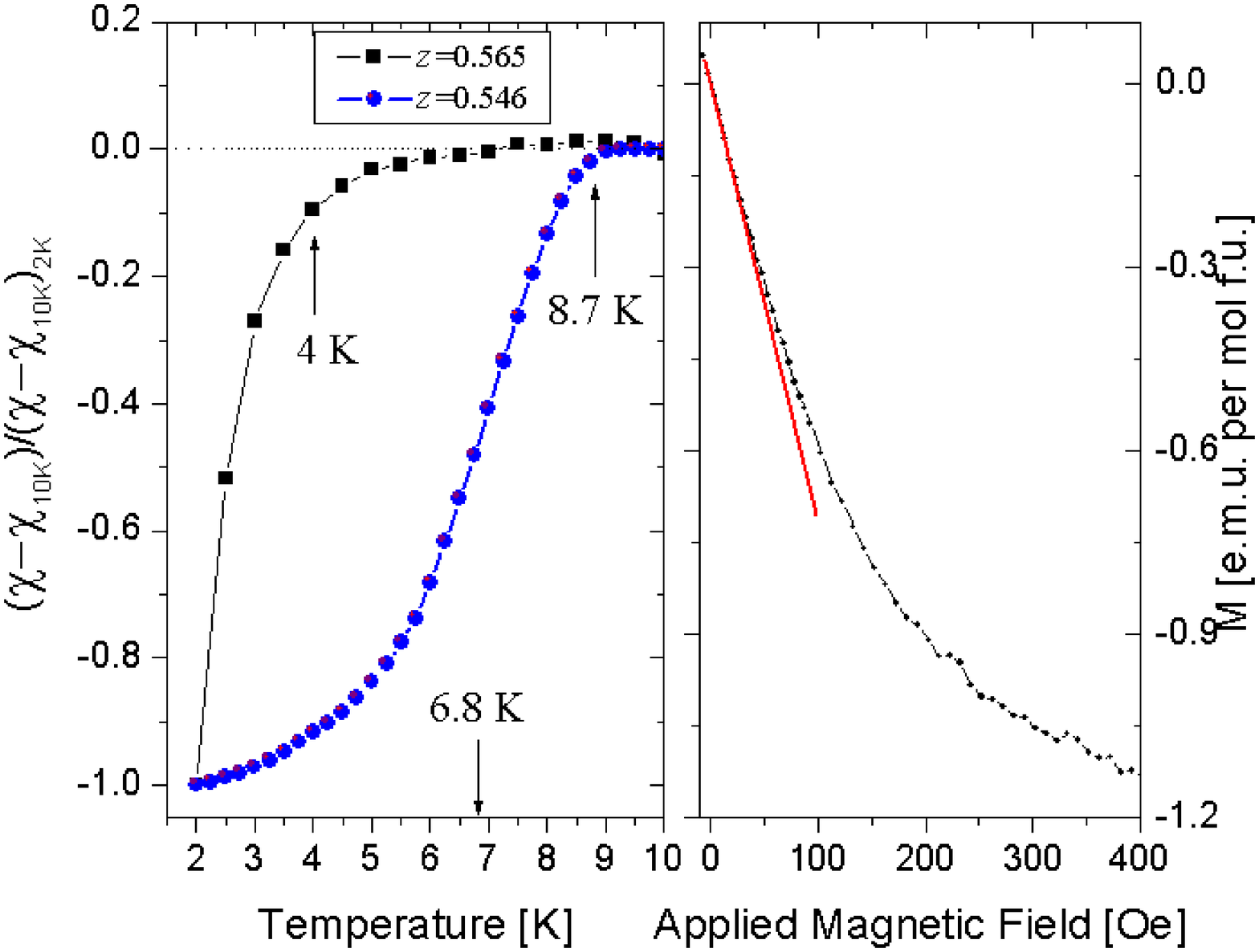}
  \caption{(Color online) Left panel: Temperature dependence of the
    ZFC dc-magnetic susceptibility measured at 20\,Oe of two trigonal
    samples of BaSi$_2$ marked by their $z$ coordinate (see text). For
    convenience, we plot the differences $\chi(T)-\chi(10K)$,
    normalized by the value at 2\,K.  The arrow indicates
    $T_\textrm{c}$ found in Ref.~\onlinecite{11}.  Right panel:
    initial magnetization curve as a function of the applied magnetic
    field measured at 2\,K in the trigonal phase with a
    $z$=0.546(3)\,\AA. The lower critical field H$_{c1}\sim$75\,Oe is
    determined from the departure from linearity (red line) at low
    field. The upper critical field H$_{c2}$ has been estimated at
    about 7 kOe by the magnetic field at which the {\it M-H} reverse
    legs merge at high magnetic field using the criteria of
    $\vert\Delta M\vert <$ 10$^{-5}$\,emu. The values of H$_{c1}$ and
    H$_{c2}$ include the demagnetization effect assumed to be 1/3 in
    cgs units.}
\label{fig:TC}
\end{figure}

Polycrystalline metastable high pressure and high temperature trigonal
samples of BaSi$_2$ were synthesized in a Belt-type apparatus from the
commercial (Cerac incorporated) orthorhombic phase of BaSi$_2$ ($Pnma$
space group) with a purity of 98\%.  Different trigonal structures
were meta-stabilized by changing the temperature and the pressure
conditions of synthesis according to the phase diagram of
BaSi$_2$~\cite{9,17,18,26}.  The structural characterizations were
performed by X-ray diffraction $\theta-2\theta$ measurements done on a
Bruker D8 Advance powder diffractometer (K$\alpha_{1,2}$ Cu
wavelengths).  The Rietveld analysis (GSAS software) of the X-ray
profiles confirms the trigonal phase with an average purity calculated
to be more than 98\%. The main phase impurity is the semi-conducting
cubic BaSi$_2$ presenting the SrSi$_2$-type structure. The three
adjustable cell parameters are the two lattice parameters $a$ and $c$
and the $z$ coordinate which represents the $2d$-Wyckoff position of
Si atoms ($z$= 0.5 corresponds to completely flat Si plans).  Among
the different samples, the $z$ coordinate exhibits sizable variations
depending on the pressure and on the temperature of synthesis. The two
samples presenting the most pronounced structural difference were
synthesized at (4.5GPa, 500$^{\circ}$\,C) and (4.5\,GPa,
1000$^{\circ}$\,C), and have, respectively, [$a$=4.061(3)\,\AA,
  $c$=5.293(3)\,\AA, and $z$=0.565(1)] and [$a$=4.065(3)\,\AA,
  $c$=5.347(4)\,\AA, and $z$= 0.546(3)]. The superconducting
transition temperatures $T_{\textrm{c}}$ of those samples were
measured in a SQUID (QD MPMS 5XL). In the left panel of
Fig.~\ref{fig:TC}, we can see that the temperature dependence of the
zero-field-cooled (ZFC) magnetic susceptibility reveals a
$T_\textrm{c}$ onset of 4\,K for the phase with $z$=0.565 and 8.7\,K
for the one with $z$=0.546. For the second sample, with $T_\textrm{c}$
close to 9\,K the {\it M-H} loop was measured at 2\,K ($z$=0.546), as
shown in the right panel of Fig.~\ref{fig:TC}. It evidences a type-II
superconducting state with H$_{c1}\sim$ 75\,Oe and H$_{c2}\sim$
7\,kOe. Using the Ginzburg-Landau equations~\cite{19}, we estimate the
penetration depth $\lambda\sim$ 3300\,\AA\ and the coherence length
$\xi=104$\,\AA.

	
Our DFT calculations were performed with the {\sc abinit}
code~\cite{20}. The exchange-correlation functional was modeled by a
Perdew, Burke, and Erzernhof (PBE) generalized gradient
approximation~\cite{21}, while the electron-ion interaction was
described by norm-conserving Troullier-Martins
pseudopotentials~\cite{22} generated with the same functional.  To
obtain the phonon dispersion we employed density-functional
perturbation theory.  Proper convergence was ensured using a cut-off
energy of 30\,Ha and a 20$^3$ $k$-mesh with the Monkhorst-Pack
sampling of the Brillouin zone. A grid of 4$^3$ $q$ phonon wave
vectors and the tetrahedron technique was used for the integration
over the Fermi surface.  In the trigonal phase the lowest energy
structure of BaSi$_2$ has lattice parameters ($a=4.08$\,\AA,
$c=5.42$\,\AA, and $z=0.56$\,\AA). These values are consistent with
our low temperature sample within the typical error of PBE
calculations.  However, by inspection of Fig.~\ref{fig:ENERGY}, we
realized that the total energy curve is rather flat in an interval of
$z$, analogously to the case of CaSi$_2$~\cite{24}.  We identify three
intervals of values of $z$: a planar region for $z \lesssim 0.545$, a
high-buckled for $z \gtrsim 0.575$, and a low-buckled for $0.545
\lesssim z \lesssim 0.575$. In the low-buckled interval, the total
energy changes by less than 50\,meV (580\,K) per unit cell, explaining
why it is easy to deform the structure, making it possible to obtain
experimental samples with different $z$.

\begin{figure}[t]
  \centering \includegraphics[width=0.85\columnwidth]{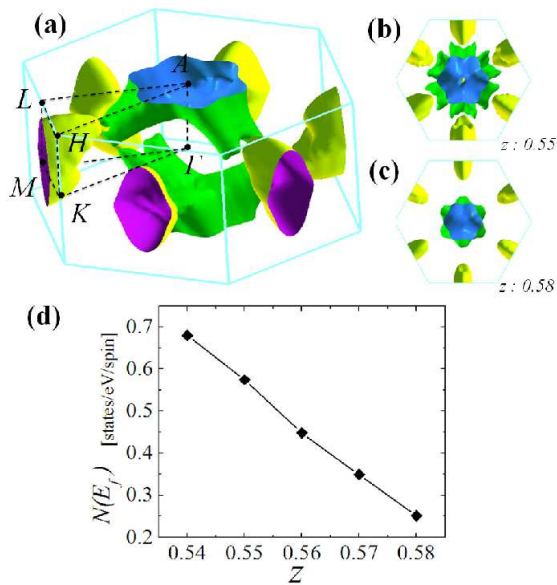}
  \caption{(Color online) Panel ({\bf a}):~Fermi surface of BaSi$_2$
    in the $P-3m1$ structure ~\cite{25}. Panels ({\bf b}) and ({\bf
      c}):~two-dimensional cuts through the 3D Fermi surface
    perpendicular to the [001] direction for $z=0.55$ and 0.58. The
    increase of the Fermi surface when the structure flattens out
    (smaller $z$) is evident. Panel ({\bf d}): Electronic
    density of states as a function of $z$.}
  \label{fig:PRE}
\end{figure}

Despite the flatness of the total energy curve, small variations of
$z$ in the low-buckled region produce remarkable changes in the
electronic structure, in the phonon modes, and in the EP coupling. The
Fermi surface, depicted in Fig.~\ref{fig:PRE}({\bf a}) is composed of
two parts, a larger portion centered at the $A$ point and smaller
pockets centered at the $M$-points. The bands composing the former
have mostly Si 3$p_{xy}$ character, and disperse much more strongly in
the $xy$ plane than along vertical directions. Increasing the buckling
of the Si-Si bond leads to a shift of these bands, which reduces the
size of the Fermi surface [panels ({\bf b}) and ({\bf c})] and,
consequently, decreases the value of the density of states at the
Fermi level [$N(E_\text{F})$ -- see panel ({\bf d})]. When one reaches
$z\gtrsim0.58$, the interacting orbitals become less confined within
the Si layers, forming a weak hybridization between the $5d$ states of
Ba and the 3$p_z$ of Si.  Also the pockets, formed by Ba $5d$ and Si
$3p_{xy}$ states, show a light decrease of area upon increase of $z$,
but on a smaller scale than the surface centered at $A$. For larger
values of $z$ (up to 0.6) the Fermi surface does not change
significantly from Fig.~\ref{fig:PRE}({\bf c}). For values of $z$
smaller than 0.55 we can see the appearance of a third sheet of the
Fermi surface composed purely of Si $3s$ states. This effect can
already be seen in Fig.~\ref{fig:PRE}({\bf b}) as a tiny yellow dot
centered at the $A$-point. Note, however, that for $z<0.545$ the
system becomes structurally unstable, making therefore this region of
limited practical importance. Looking at the density of states at the
Fermi level [Fig.~\ref{fig:PRE}({\bf d})] we see that it attains a
maximum in the low-buckled region ($z=0.545$), then it decreases with
increasing $z$. These numbers already suggest that, from the
electronic point of view, superconductivity is favored by smaller
values of $z$.


\begin{figure}[t]
  \centering \includegraphics[width=1.05\columnwidth]{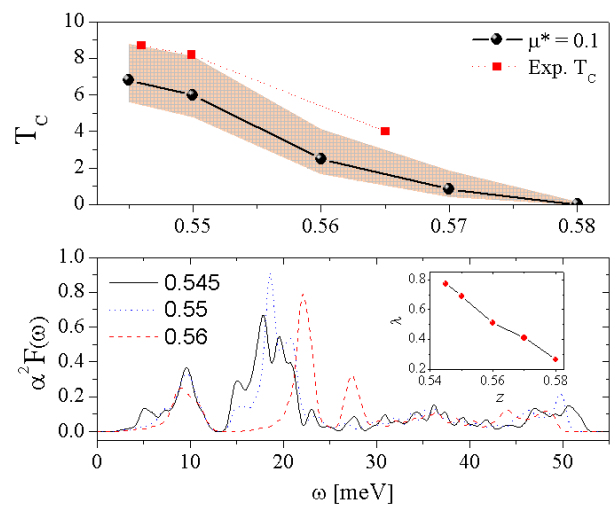}
  \caption{(Color online) Bottom panel: Eliashberg spectral function
    $\alpha^2F(\omega$) (in the inset: EP coupling constant $\lambda$)
    as a function of $z$. Top panel: $T_\text{c}$ in K calculated with the
    McMillan formula. The shaded area shows the interval of
    $T_\text{c}$ values when $\mu^*$ varies from 0.06 to 0.125.  For
    comparison the dashed line represent the experimental values of
    $T_{\textrm c}$. }
  \label{fig:LAM}
\end{figure}

In order to verify this point, we calculated both the phonon modes and
the EP coupling constants as a function of $z$. At the equilibrium
structure our calculated phonon frequencies are close to experimental
values~\cite{26}. For $z$ close to 0.5, \textit{i.e.} for almost
planar Si layers, the modes exhibit imaginary frequencies indicating a
structural instability. In the stable region, most phonon frequencies
show only a weak dependence on buckling. However, the $A_{1g}$ optical
phonons, which are mainly composed of vibrations of the Si atoms along
the $c$ axis (buckling modes), considerably soften upon decreasing
$z$, leading eventually to the structural instability. Moreover, the
higher energy $E_g$ mode, composed of Si vibrations in the $xy$
planes, is slightly hardened with decreasing $z$. In spite of these
changes, the value of $\Omega_\text{log}$, the weighted average of the
phonon frequencies, remains basically unchanged in the optimal range
$0.55-0.58$, and therefore can not be responsible for the variations
of $T_\text{c}$. At $z=0.56$ we obtain the maximal value
$\Omega_\text{log}=182$\,K; at $z=0.55$ it is slightly smaller,
$\Omega_\text{log}$=170\,K, for $z=0.545$ it drops to 156\,K and for
$z=0.57$ and $z=0.58$ we found $\Omega_\text{log}=179$\,K and 171\,K,
respectively.


To evaluate the superconducting transition temperature $T_\text{c}$ in
the framework of the strong coupling theory of
superconductivity~\cite{27,28} we need to calculate $\lambda$, which
measures the average EP interaction (inset of the bottom panel of
Fig.~\ref{fig:LAM}). As the calculation of $\lambda$ involves
averaging over the Fermi surface, special care has to be taken to
ensure the convergence of the results. In fact, an insufficient
sampling can lead to a dramatic underestimation of
$T_\text{c}$~\cite{12}. The screened coulomb interaction $\mu^*$ was
set to 0.1, which is a standard choice for this kind of material.  The
Eliashberg spectral function $\alpha^2F(\omega)$ is shown in the
bottom panel of Fig.~\ref{fig:LAM}. There are two main peaks
contributing to $\lambda$: the first, due to the acoustic modes, is
fairly insensitive to changes in $z$; the second peak, due to the EP
coupling of the $A_{1g}$ optical modes, moves to lower frequency,
increasing its area and consequently its contribution to $\lambda$
with decreasing $z$. In the inset of the bottom panel of
Fig.~\ref{fig:LAM} one can see how this leads to a dramatic increase
of the EP coupling constant $\lambda$ with $z$, and therefore to a
maximum theoretical $T_\text{c}$ of around 6\,K using the McMillian
formulation~\cite{29} (top panel of Fig.~\ref{fig:LAM}).  Note that
the experimental trend for the dependence of $T_\text{c}$ of BaSi$_2$
on the buckling is perfectly reproduced by our calculations, which
explain also previous experimental findings for the similar compound
CaSi$_2$.  The slight underestimation of calculated $T_\text{c}$ is
due to the neglect of multi-band effects that are known to enhance
superconductivity in similar systems, like MgB$_2$~\cite{4,6}.


In conclusion, we presented an experimental and theoretical study of
the layered binary silicide BaSi$_2$. We show that the buckling of the
Si sheets can be modified experimentally by using different
high-pressure and temperature conditions of synthesis. The reason for
such a behavior can be found in our DFT calculations, which evidence a
broad low-buckling interval where the total energy changes by less
than 50\,meV (580\,K) per unit cell as a function of $z$.  The
electronic band structure calculations demonstrate that the density of
states at the Fermi level significantly increases by reducing the
buckling of the Si planes.  However, the flattening of the Si layers
is limited by a structural instability concomitant with the softening
of the Si optical buckling phonon mode (A$_{1g}$).  The coupling of
this mode with electrons is also dramatically enhanced by flattening
the Si planes, leading to the doubling (from 4 \,K up to 8.7\,K) of
the superconducting transition temperature. Such competition between
superconductivity and structural distortion can be found in a wide
variety of conventional superconductors, like the Chevrel-phases, the
transition metal carbides, the cubic Ba$_{1-x}$K$_x$BiO$_3$, etc.
This scenario is in full agreement with our experimental and
theoretical findings of an increase of $T_\text{c}$ when Si planes
flatten out, and it is compatible with previous measurements on the
disilicide CaSi$_2$ ~\cite{15,16}. Moreover, the mechanism of tuning
$T_\text{c}$ by controlling the buckling of the layers is likely to be
present in many other layered superconductors, and therefore can
provide a new path for optimizing $T_\text{c}$.


\begin{acknowledgements}
SB acknowledges support from EU’s 7th Framework Programme (e-I3
contract ETSF) and MALM from the French ANR
(ANR-08-CEXC8-008-01). Calculations were performed at GENCI (project
x2011096017). JAFL acknowledges the program CONACyT-Mexico. We thank
D. Machon, R. Viennois, D. Ravot, P. Toulemonde and Y. Sidis for
discussions and S. Le Floch for sample preparation.  We acknowledge
V. Dupuis, M. Hillenkamp and D. Luneau from the Centre de
Magn\'{e}tom\`{e}trie of Lyon for the use of the Squid and R. Vera and
E. Jeanneau from the Centre de Diffractom\`{e}trie Henri Longchambon
for the X-Ray measurements at the University of Lyon 1.
\end{acknowledgements}


\begin{thebibliography}{ll}
\bibitem{19} $\mu_0H_{c2}=\Phi_0/2\pi\epsilon^2$ and $\mu_0 H_{c1}=(\Phi_0 / 4\pi\lambda^2)[ln(\lambda/\epsilon)+C1]$ where $\Phi_0$ is the flux quantum and C1=0.497.
\bibitem{26} Details on the synthesis procedure and the Raman studies will be published elsewhere by the same authors.
\end{thebibliography}
\end{document}